\documentclass[11pt]{article}

\title{\bf Radiative transfer in a spherical, emitting, absorbing and anisotropically scattering medium}
\author{Michael J. Caola\\6 Normanton Rd., Bristol, BS8 2TY, UK\\ caola@blueyonder.co.uk}
\usepackage{graphics}
\usepackage{amsmath}
\begin{document}
\maketitle
\begin{abstract}
\noindent
The atmospheres of planets (including Earth) and the outer layers of stars have often been treated in radiative transfer as plane-parallel media, instead of spherical shells, which can lead to inaccuracy, e.g. limb darkening. We give an exact solution of the radiative transfer specific intensity at all points and directions in a finite spherical medium having arbitrary radial spectral distribution of: source (temperature), absorption, emission and anisotropic scattering. The power and efficiency of the method stems from the spherical numerical gridding used to discretize the transfer equations prior to matrix solution: the wanted ray and the rays which scatter into it both have the same physico-geometric structure. Very good agreement is found with  an isotropic astrophysical benchmark (Avrett \& Loeser, 1984). We introduce a specimen arbitrary forward-back-side phase scattering function for future comparisons. Our method directly and exactly addresses spherical symmetry with anisotropic scattering, and could be used to study the Earth's climate, nuclear power (neutron diffusion) and the astrophysics of stars and planets.                                                    
\end{abstract}

\section{Introduction}
Radiative transfer grew largely from astrophysics [1,2] and later moved to terrestrial problems of atmosphere [3], hot gas and solids [4,5] and nuclear power [6].  Many of these problems dealt with the outer layers of a spherical medium  but, for historical, technical and other reasons, these were treated as plane parallel media [1], and the theoretical results were often very satisfactory. Nevertheless, there presently appears to be no {\it ab initio} direct systematic accurate accepted and used method for calculating radiative transfer in a  a spherical, emitting, absorbing and {\it anisotropically} scattering medium. We now present such a calculation, but firstly recall the `general' transfer equation (RTE):
$$
\frac{dI(\bf r,s)}{ds}=-I(\bf r,s)[\alpha(\bf r,s)+\sigma(\bf
  r,s)]+\alpha(\bf {r,s})\it{B}(\bf {r,s})
$$
\begin{equation}
+\frac{\sigma(\bf r,s)}{\rm{4}\pi}\int_{\rm{4}\pi}\it{I'}(\bf r,s')\it{F}(\bf r,s,s')\it{d\Omega}(\bf r,s')
\end{equation}
\par
Here $I(\bf r,s)$ is the (specific) intensity of photons inside the medium at point $\bf r$ and detector direction $\bf s$, $B(\bf r,s)$ is the photon source function, $\alpha(\bf r,s)$ is the absorption coefficient, $\sigma(\bf r,s)$ is the scattering coefficient, and a fraction (part)  $F(\bf r,s,s')\it{d\Omega}(\bf r,s')$ of the non-detector intensity $I'(\bf r,s')$ is scattered into the detector direction  $I(\bf r,s)$; the scattering phase function is  $F(\bf r,s,s')$ and   $d\Omega(\bf r,s')$ the solid angle around direction $\bf s'$.
\par
Before analysing spherical symmetry, we note that (1) has the form of a many-particle transport equation: radiative transport for photons [2,3], neutron diffusion for nuclear reactions [6] and the Boltzmann equation for gas molecules [7].

\section{Analysis}
We calculate  the intensity  by numerical solution of the of the
sphere gridded as in Fig.1. Spherical shells $m=1 .. M$ between
adjacent radii $r_m$ contain abritrary constant values of source $B_{m}$, absorption $\alpha_{m}$ and scattering $\sigma_{m}$. With mid-radii
\begin{equation}
\bar{r}_{m}=\tfrac{1}{2}(r_{m}+r_{m+1}),
\end{equation}
we define the intensity $I_{m,j}$ as a ray tangential to $\bar{r}_{m}$ with $j$ shells to the left.
 We introduce the standard {\it albedo} $\omega=\sigma/(\sigma+\alpha)$ and incremental {\it optical depth}
\begin{equation}
\Delta\tau_{m,j}=\Delta s_{m,j}(\alpha_{m,j}+\sigma_{m,j}),
\end{equation}  
where $\Delta s_{m,j}=ab$ is the physical length step $j$ along $I_{m,j}$. Then the general (1) becomes
\begin{equation}
\frac{\Delta I_{m,j}}{\Delta\tau_{m,j}}=-I_{m,j}+(1-\omega_{m,j})B_{m,j}+\frac{\omega_{m,j}}{4\pi}\sum_{m'=1}^{mm}\sum_{j'=1}^{2m'-1}\log_{m,j}^{m',j'}I'_{m',j'}F_{m'}^{m,j}\Delta\Omega_{m',j'}
\end{equation}
The {\it logic} term $\log{_{m,j}^{m',j'}}$ says which of the many
$I'_{m',j'}$ physically scatter at and along $I_{m,j}$ and, together with step $\Delta s_{m,j}$, phase $ F_{m'}^{m,j}$ and solid-angle $\Delta\Omega_{m',j'}$, are derived in the Appendix, all as functions of the the shell radii $r_{m}$.
\par
We now make an essential observation that in (4) the set of detector
intensities $\{I_{m,j}\}$ is identical to the the set of scattering
intensities  $\{I'_{m',j'}\}$: this is visually obvious from Figure 1(a,b), and
would not necessarily be so had we chosen a different geometric
gridding. Thus $I=I'$, and with $\Delta I_{m,j}=I_{m,j}-I_{m,j-1}$, we
can simply rearrange (4) as
$$
I_{m,j}=\frac{B_{m,j}\Delta\tau_{m,j}(1-\omega_{m,j})}{1+\Delta\tau_{m,j}}+
$$ 
\begin{equation}
\sum_{m'=1}^{mm}\sum_{j'=1}^{2m'-1}\frac{\delta_{m',m}\delta_{j',j-1}
  +\frac{\Delta\Omega_{m',j'}}{4\pi}\Delta\tau_{m,j}F_{m'}^{m,j}\log_{m,j}^{m',j'}\omega_{m,j}}{1+\Delta\tau_{m,j}}I_{m',j'}
\end{equation}

In (5) we can count the integer index pairs $(m,j)$ and $(m',j')$ by
single integers $i$ and $i'=1 .. M^2$:
$$
i=i(m,j)=(m-1)^2+j
$$
\begin{equation}
i'=i(m',j')=(m'-1)^2+j'
\end{equation}
and (5) becomes
\begin{equation}
I_{i}=S_{i}+\sum_{i'}K_{i,i'}I_{i'} 
\end{equation}
where the source vector $S$ is
\begin{equation}
S_{i}=S_{i(m,j)}=\frac{B_{m,j}\Delta\tau_{m,j}(1-\omega_{m,j})}{1+\Delta\tau_{m,j}}
\end{equation}
and the scattering kernel matrix $K$ is
\begin{equation}
K_{i,i'}=K_{i(m,j),i(m',j')}=\frac{\delta_{m',m}\delta_{j',j-1}+\frac{\Delta\Omega_{m',j'}}{4\pi}\Delta\tau_{m,j}F_{m'}^{m,j}\log_{m,j}^{m',j'}\omega_{m,j}}{1+\Delta\tau_{m,j}}.
\end{equation}
The matrix equation (7), $I=S+KI$, has solution $I=(1-K)^{-1}S$ given
by a standard numerical linear algebra solver, e.g. BLAS. An
$I_{i}=I_{i(m,j)}=I_{m,j}$ is the intensity at any point with radius
$\bar{r}_{m-|m-j|}$ and at angle $\sin^{-1}(\bar{r}_{m}/\bar{r}_{m-|m-j|})$ to
it. Matrix solution $I$ of (7) gives $M^2$ exact values of $I_{m,j}$,
whether or not we need them all. The $I_{m,2m-1}$ are {\it surface}
intensities, as detected {\it in vacuo} at distance $R>r_{1}$, the
others are internal intensities. 

\section{Results}
We calculate the astrophysical benchmark of Avrett \& Loeser '84 [8] by our method. Their results are the basis of other comparisons [9-14]. We have implemented our analysis {\S}2 as computer code in both Fortran and Mathcad, giving identical numerical results. The results are for the directionally averaged intensity $J_m$ in shell $m$,
\begin{equation}
J_m=\frac{1}{4\pi}\sum_{m'=1}^{M}\sum_{j'=1}^{2m'-1}\delta(m,m'-|m'-j'|)[1+\delta(m',j')]I_{m'j'}\Delta\Omega_{m,j}
\end{equation}

See [8] for details of the code inputs \{$M,r_m,B_m,\alpha_m, \sigma_m,F=1$\}, isotropic scattering. Our results are in (very) good agreement with [8], see Fig. 2, which attest to the validity and value of our method.

\section{Discussion and applications}
\begin{itemize}
\item
All quantities so far are implicictly monochromatic, at photon frequency $\nu$. If shell $m'$ is in local thermodynamic equilibrium (LTE) then it may be assigned a temperature $T_{m'}$, and the source function is the Planck black-body function:
%\begin{equation}
$$
B_{m,j}=B_{m,j}^{(\nu)}=\frac{k_{1}\nu^{3}}{\exp(h\nu/kT_{m-|m-j|})-1}
$$
%\end{equation}

\item
The ratio $f$:$s$:$b$ determines the forward, side and back strength of the anisotropic scattering {\it eglipsoid}, see the Appendix. We have calculated the intensity $I_{m,j}$  in an homogenous $B=1$ $\omega=0.5$ $R=1$  $M=5$ shell sphere for the cases (f,s,b) = (2,2,2), (1,2,3), (1,3,2), (2,1,3), (2,3,1), (3,2,1), (3,1,2). These give a quantitative feel for the effects of anisotropic scattering, and are available on request from the author.
 
\item  
We can simulate a planetary atmosphere (e.g. Earth) by $h$ shells $r_1..r_h$ as gas and shells $r_{h+1} .. M$ as solid, with shell/surface $r_h$ having emissivity  $\epsilon$ (and thus reflectivity $1-\epsilon$). A planetary atmosphere is typically `thin', $r_1-r_h\ll r_M$, and has `high' absorption $\alpha_{solid}\gg\alpha_{gas}$ in the solid core. Rays $I_{m,j}$ with $m<h$ are `limb' rays which do not intercept the solid planet and are not easily modelled in a plane parallel situation, but are rays like any others for our method. If the planet receives the Sun's intensity $I_{solar}$ from left to right, then we have boundary conditions
\begin{equation} 
I_{m,j}=I_{m,j=0}=I_{m,0}=I_{solar},
\end{equation}
with day in the left hemisphere and night in the right.\\
\\
Realistic code implementation of the the above planetary physics, especially our calculation with a total planetary volume mainly of high-$\alpha$ solid and its surface, needs careful comparison to present methods to judge if the differences between our `direct-sphere' method and conventional methods (plane-parallel) are of value and advantage.  
\end{itemize}

\section*{Appendix}
We here give expressions for $\log_{m,j}^{m',j'}$, $\Delta s_{m,j}$,
$\Delta\Omega_{m',j'}$ and $F_{m'}^{m,j}$; all derive from elementary
spherical/circular geometry.\\
\\
{\bf Scattering `selection rules'.} From Figure 1(b) we see that the
detector ray $I_{m,j}$ is on shell number 
\begin{equation}
m_{j}=m-|m-j| \qquad m_{j}=m_{2m-j}
\end{equation}
 and thus only (scattering) rays $I_{m',j'}$ with $m'_{j'}=m_{j}$ can
 scatter into it. With (10) this means
\begin{equation}
\log_{m,j}^{m',j'}=\delta(m'_{j'},m_{j})[1+ \delta(j',m')]
\end{equation}\\
\\
{\bf Physical step length.} From Figure 1(c), this is $ab=$
\begin{equation}
\Delta s_{m,j}=
\left\{\begin{array}{cl}
\sqrt{r_{m-|m-j|}^{2}-\bar{r}^{2}_{m}}-\sqrt{r_{m-|m-j-1|}^{2}-\bar{r}^{2}_{m}}&
 \mbox{if
  $0<j<m$}\\
2\sqrt{r_{m}^{2}-\bar{r}^{2}_{m}}& \mbox{if $j=m$}\\
\Delta s_{m,2m-j} & \mbox{if $2m>j>m$}
\end{array}\right.\
\end{equation}\\
\\
{\bf Solid angle.} A cone of semi-angle $\theta$ subtends solid angle
$\Omega(\theta)=2\pi(1-\cos\theta)$ and, see Figure 1(c) which is a central circular section of the sphere, ray $I_{m,j}$ is on the cone $\Omega(AjA')$  whose surface A$j$ is swept out by revolution about axis O$j$, and contains the tangent sphere radius $\bar{r}_{m} $: the solid angle of $I_{m,j}$ in (4) is then the conical shell between adjacent cones $\Omega(BjB')$ and $\Omega(CjC')$, containing tangent spheres $r_{m}$ and $r_{m+1}$:
$$
\Delta\Omega_{m,j}=\Omega(BjB')-\Omega(CjC')
$$

\begin{equation}
=2\pi
\left\{\begin{array}{cl}
1 &\mbox{if $m=M$ and $j=m$}\\
1-\sqrt{1-r_{m+1}^{2}/\bar{r}^{2}_{m-|m-j|}}   &\mbox{if $m=M$ and $j\ne m$}\\
\sqrt{1-r_{m}^{2}/\bar{r}^{2}_{m-|m-j|}} &\mbox{if $m< M$ and $j=m$}\\
\sqrt{1-r_{m+1}^{2}/\bar{r}^{2}_{m-|m-j|}}-\sqrt{1-r_{m}^{2}/\bar{r}^{2}_{m-|m-j|}} &\mbox{if $m<M$ and $j\ne m$}
\end{array}\right.\
\end{equation}\\
\\
{\bf Scattering phase function.} The analysis so far directly covers isotropic scattering, $F=1$, and we now turn to the detail of anisotropic scattering.   For most scattering media of interest
  (random, homogenous, not `oriented' or crystalline) the strength of anisotropic
  scattering depends only on the angle $\gamma=\gamma_{p,m,j}^{p',m'}$ between the
  detector ray $I_{p,m,j}$ and the scattering ray $I_{p',m',j'}$; in Fig. 1(c) with polar axis $z=jO$ and $x$ on the paper surface, $\phi_{p}=2\pi(p-1)/P$ is the azimuthal angle of  $I_{p,m,j}$ on the detector cone $m$ and likewise   $\phi_{p'}$ is the azimuthal angle of  $I_{p',m',j'}$ on the scattering cone $m'$. Then, with $\sin\theta_{m,j}=\bar{r}_{m}/\bar{r}_{m-|m-j|}$,
\begin{equation}
\cos\gamma_{p,m,j}^{p',m'}=\cos\theta_{m,j}\cos\theta_{m',j}+\sin\theta_{m,j}\sin\theta_{m',j}\cos(\phi_{p}-\phi_{p'})
\end{equation}
Since symmetry says the {\it magnitude} of intensity $I_{p,m,j}$ is independent of $\phi_{p}$, we choose  $\phi_{p}=0$ and can thus write   $I_{p,m,j}=I_{m,j}$ and $\cos\gamma_{p,m,j}^{p',m'}=\cos\gamma_{m,j}^{p',m'}$.
\\\\
The scattering strength is given by some function $F'(\gamma)$ defining a non-spherical surface. For illustration we choose surface $F'$ to be egg-shaped, being the smooth union of two different half-ellipsoids of revolution, an {\it eglipsoid},  defined by forward $f$, side $s$ and back $b$ major/minor axes:
\begin{equation}
F'(v)=
\left\{\begin{array}{cl}
N(f,b,s)/\sqrt{v^{2}f^{-2}+(1-v^{2})s^{-2}}     &\mbox{if $1\ge v \ge 0$}\\
N(f,b,s)/\sqrt{v^{2}b^{-2}+(1-v^{2})s^{-2}}     &\mbox{if $-1\le v < 0$}\\
\end{array}\right.\
\end{equation}
where $N$ ensures phase normalisation, $\int F'd\Omega=4\pi$. Thus a fraction $F'(\gamma_{m,j}^{p',m'})$ of $I_{p',m',j'}$ is scattered along $I_{m,j}$ (=$I_{p,m,j}$), and phase function $F^{m,j}_{m'}$ is the sum of all $p'=1,2, .. P$ such contributions:
\begin{equation}
 F^{m,j}_{m'}=\frac{1}{P}\sum_{p'=1}^{P}F'(\cos\gamma_{m,j}^{p',m'})
\end{equation}

\pagestyle{empty}
\begin{figure}
\thispagestyle{empty}
%\begin{centre}
%\resizebox{60mm}{!}{}
%%\includegraphics{grid.eps}
\caption{Geometric spherical gridding of our analysis, {\S}2; (a) and (b) show that the set $\{I\}$ of detector rays is identical to the scattering set  $\{I'\}$, and (c) shows constructions for step $\Delta s_{m,j}$, solid angle $\Delta\Omega_{m,j}$ and polar axes $xyz$ for the azimuthal $\phi_{p}$ of anisotropic scattering.}
%\end{center}
\end{figure}

\begin{figure}
\thispagestyle{empty}
%\begin{centre}
%\resizebox{60mm}{!}{}
%%\includegraphics{av84mc06.eps}
\caption{Comparison of benchmark [8] (circles), to our results (lines) {\S}3. The $M=$14 shell sphere has optical density (depth) $\tau\propto 1/r$. }
%\end{center}
\end{figure}

\end{document}